\documentclass[aps,prb,preprint,superscriptaddress]{revtex4-1}
\usepackage{graphicx}
\usepackage{dcolumn}
\usepackage{bm}
\usepackage{color}
\usepackage{array}

\begin{document}


\title{Temperature- and pressure-dependent metallic states in
(BEDT-TTF)$_8$[Hg$_4$Br$_{12}$(C$_6$H$_5$Br)$_2$]}



\author{Alain~Audouard}
\email[E-mail address: ]{alain.audouard@lncmi.cnrs.fr}
\affiliation{Laboratoire National des Champs Magn\'{e}tiques
Intenses (UPR 3228 CNRS, INSA, UJF, UPS) 143 avenue de Rangueil,
F-31400 Toulouse, France.}

\author{Fabienne~Duc}
 \affiliation{Laboratoire National des Champs Magn\'{e}tiques
Intenses (UPR 3228 CNRS, INSA, UJF, UPS) 143 avenue de Rangueil,
F-31400 Toulouse, France.
}%

\author{David~Vignolles}
\affiliation{Laboratoire National des Champs Magn\'{e}tiques
Intenses (UPR 3228 CNRS, INSA, UJF, UPS) 143 avenue de Rangueil,
F-31400 Toulouse, France.}

\author{Rustem~B.~Lyubovskii}
\affiliation{Institute of Problems of
Chemical Physics, RAS, 142432 Chernogolovka, MD, Russia.}

\author{Laure~Vendier}
 \affiliation{CNRS, LCC (Laboratoire de Chimie de Coordination), 205 route de Narbonne, F-31077 Toulouse, France.
}%

\author{Gena~V.~Shilov}
 \affiliation{Institute of Problems of
Chemical Physics, RAS, 142432 Chernogolovka, MD, Russia.
}%

\author{Elena~I.~Zhilyaeva}
 \affiliation{Institute of Problems of
Chemical Physics, RAS, 142432 Chernogolovka, MD, Russia.
}%

\author{Rimma~N.~Lyubovskaya}
 \affiliation{Institute of Problems of
Chemical Physics, RAS, 142432 Chernogolovka, MD, Russia.
}%

\author{Enric~Canadell}
 \affiliation{Institut de Ci\`{e}ncia de Materials de Barcelona, CSIC, Campus de
la UAB, 08193, Bellaterra, Spain.
}%


\date{\today}

\begin{abstract}
Temperature-driven metal-insulator and pressure-driven insulator-metal transitions observed in (BEDT-TTF)$_8$[Hg$_4$X$_{12}$(C$_6$H$_5$Y)$_2$] with X = Y = Br are studied through band structure calculations based on X-ray crystal structure determination and Shubnikov-de Haas (SdH) oscillations spectra, respectively. In connection with chemical pressure effect, the transition, which is not observed for X = Cl, is due to gap opening linked to structural changes as the temperature decreases. Even though many body interactions can be inferred from the pressure dependence of the SdH oscillations spectra, all the data can be described within a Fermi liquid picture.
\end{abstract}

\pacs{61.50.Ks, 71.20.Rv, 71.18.+y, 71.30.+h  }

\maketitle

\section{\label{sec:Intro}Introduction}

\begin{figure*}
\includegraphics[width=1.0\columnwidth,clip,angle=0]{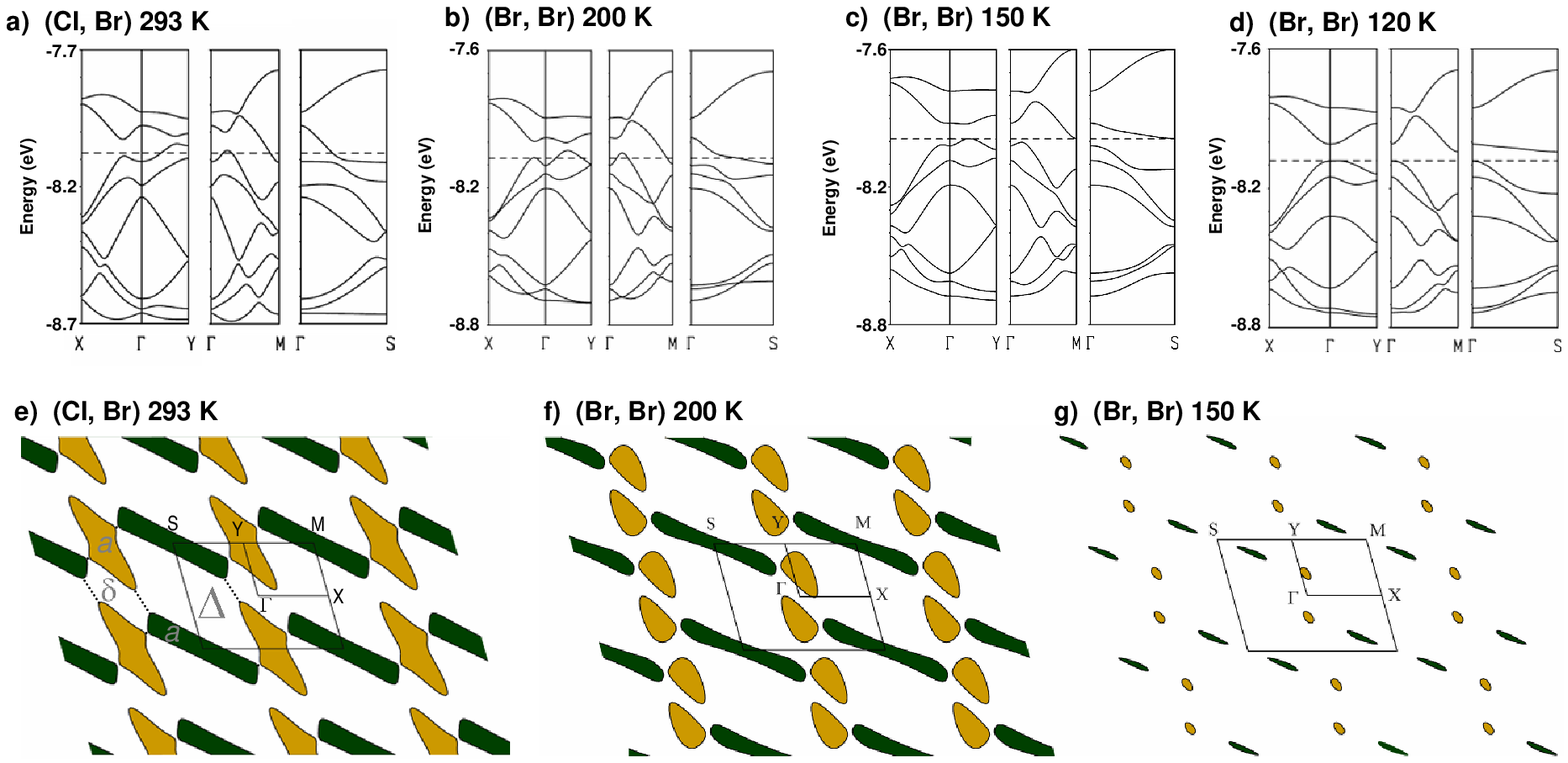}
\caption{\label{Fig:BS_FS} (color online) Band structure of (a) (Cl, Br) at 293 K, (b) to (d) (Br, Br) at 200 K, 150 K and 120 K, respectively. Dashed lines mark the Fermi level in (a), (b) and (c) and the top of the HOMO band in (d). Corresponding Fermi surfaces of (e) (Cl, Br) at 293 K, (f) and (g) (Br, Br) at 200 K and 150 K, respectively. Dark green [dark grey] and tan [light grey] areas stand for electron and hole orbits, respectively. Dashed lines in (e) mark the magnetic breakdown gaps; labels $a$ depict the closed electron and hole orbits (arrows indicate the quasi particles path) which have the same area; $\delta$ and $\Delta$ are 'forbidden orbits' liable to enter magnetic breakdown orbits. Data for (Cl, Br) are from Ref.~\onlinecite{Vi08b}.}
\end{figure*}

According to room temperature X-ray diffraction data \cite{Ly91,Dy91,Dy96}, the four charge transfer salts (BEDT-TTF)$_8$[Hg$_4$X$_{12}$(C$_6$H$_5$Y)$_2$] (where X, Y = Cl, Br and BEDT-TTF stands for bis-ethylenedithio-tetrathiafulvalene) are isostructural compounds. In the following, they are referred to as (X, Y). Even though a metallic ground-state is observed for both (Cl, Cl) and (Cl, Br), i.e. X = Cl, a metal-insulator transition is reported as the temperature is lowered for X = Br.\cite{Ly93} Band structure calculations yield Fermi surface (FS) composed of one electron and one hole compensated quasi-two-dimensional orbit for both (Cl, Cl) \cite{Ve94} and (Cl, Br)  \cite{Vi08b}, referred to as the $a$ orbits (see Fig.~\ref{Fig:BS_FS}). Quantum oscillations spectra of these compounds yield frequencies F$_a$ = 241.5 $\pm$ 2.0 T for (Cl, Cl) \cite{Pr02} and 235.5 $\pm$ 1.0 T for (Cl, Br) \cite{Vi03,Au05,Vi08b} which corresponds to orbits area of 11\% and 16\% of the first Brillouin zone (FBZ) area, respectively, in satisfactory agreement with band structure calculations. In large enough magnetic field, the orbits are connected by magnetic breakdown (MB) junctions. As a result, these FS can be regarded as textbook cases for the study of quantum oscillations in networks of compensated orbits.\cite{Fo08} In line with this statement, numerous Fourier components that are linear combinations of F$_a$ and of the frequencies linked to the $\delta$ and $\Delta$ pieces (see Fig.~\ref{Fig:BS_FS}) are observed in the quantum oscillations spectra of these two compounds.

\begin{table}
  \caption[Refinement]{\label{tab.crystdata}Experimental data
  and agreement factors for the structure refinements of (BEDT-TTF)$_8$[Hg$_4$Br$_{12}$(C$_6$H$_5$Br)$_2$] at 100, 120,
  150, 200 and 293 K. Reflections with I $>$ 3$\sigma$(I) were considered as observed ($\sigma$(I) corresponds to the
  estimated standard deviations of measured individual intensities I).}
  \begin{tabular}{l c c c c c}
    \hline
    Temperature (K) & $100$ & $120$ & $150$ & $200$& $293$ \tabularnewline

    \hline
    Maximum $\theta$ (deg)& 32 & 29 & 32 & 29 & 32\tabularnewline
    Redundancy &1.65 & 2.81& 1.67 &2.83 & 1.70\tabularnewline
    Number of measured reflections & $18078/37105$ & $24105/48089$& $18582/38435$ & $19996/48673$ &$13408/40197$ \tabularnewline
     (observed/all) &  &  & &  &   \tabularnewline
    Unique reflections (observed/all) & $10560/22447$ & $9025/17107$ & $10796/22940$ & $7345/17186$ & $7151/23540$  \tabularnewline
       Rint (observed/all)       & $6.82/8.79$ & $3.98/4.89$ & $3.96/4.97$ & $4.15/5.70$ &$6.09/9.00$ \tabularnewline
    Rwobs/Rwall       & $8.75/9.14$ & $2.93/3.10$  & $5.91/6.13$ & $3.76/4.02$ & $8.81/8.89$\tabularnewline
    g.o.f.obs/ g.o.f.all       &  $3.57/2.53$& $1.16/0.88$&$2.40/1.69 $& $1.47/1.01$ &$3.71/2.11$\tabularnewline
    Number of refined parameters & $554$ & $554$ & $554$ & $554$ & $554$\tabularnewline
    \hline
  \end{tabular}
\end{table}

 Oppositely, both (Br, Cl) and (Br, Br) have an insulating ground state.\cite{Ly93} Nevertheless, a metallic ground state is recovered for (Br, Br) under an applied pressure of 0.4 GPa and Shubnikov-de Haas (SdH) oscillations have been observed at 0.8 GPa in fields below 15 T.\cite{Ly09} Their frequency (F = 215 T) is close to that linked to the closed orbits $a$ of the compounds with X = Cl, i.e. with metallic ground state at ambient pressure. This could suggest that the pressure-induced metallic state in (Br, Br) is similar to the ambient pressure ground state of (Cl, Cl) and (Cl, Br) and offer the possibility to study the applied pressure-induced evolution of the ground state. In that respect, metal-insulator transition in strongly correlated fermions systems have received much attention. As an external parameter such as doping or applied pressure is varied, numerous phases such as Mott insulator, superconducting and density wave states, eventually including quantum critical points, are observed. As examples, quantum critical points have been inferred from doping dependence of the effective mass in both superconducting cuprates \cite{Si10} and iron pnictides.\cite{Sh10}

In this work, we first focus on the temperature dependence of atomic and electronic structures of (Br, Br) in order to get information about the metal-insulator transition. It is concluded that the insulating state at ambient pressure is related to progressive band separation linked to chemical pressure as the temperature decreases, followed by a subtle conformational change of two of the donors in the vicinity of the transition, leading to gap opening. In a second step, the pressure-induced metallic state is considered. The FS topology deduced from SdH oscillations spectra is different from the FS in the metallic state at room temperature but similar to that of the compounds with X = Cl, indicating that applied pressure and temperature increase have different effects. Finally, even though an increase of many body effects is observed as approaching the insulating state as the applied pressure decreases, the data are in line with a Fermi liquid picture throughout the pressure range explored.

\section{\label{sec:expt}Experimental}

The crystals studied were synthesized by the
electrocrystallization technique reported in Ref. \onlinecite{Ly93}.
Their mean dimensions were {0.1 $\times$ 0.1 $\times$ 0.02}
mm$^3$ for X-ray diffraction experiments and 0.5 $\times$ 0.5 $\times$ 0.02 mm$^3$
for magnetoresistance measurements.

Single crystal X-ray diffraction data were collected at selected
temperatures (100, 120, 150, 200, 293 K) with an Oxford
Diffraction Xcalibur CCD diffractometer equipped with a Cryojet cooler device from Oxford Instruments and using a graphite-monochromated Mo-K$\alpha$
radiation source ($\lambda$ = 0.71073 {\AA}).
The unit cell determination and data reduction were carried out using the CrysAlisPro package.\cite{Crys10}
Absorption correction was performed using the multiscan procedure. All the structures were solved by direct methods using SIR92 \cite{Alt93} and refined with the JANA2006 program.\cite{Pe06}
All atoms except carbon were refined anisotropically. The hydrogen atoms were not placed in either structure.
Experimental details on the crystallographic data and structure refinements are given in Table~\ref{tab.crystdata}.
Besides the complete data collections described above, the temperature dependence of cell parameters was obtained by
refining the position of 250 to 400 reflections
between 100 and 300 K with temperature steps in the range 10 to 25 K.

Tight-binding band structure calculations were based upon the
effective one-electron Hamiltonian of the extended H\"{u}ckel
method.\cite{Wh78} A modified Wolfsberg-Helmholtz formula was used to calculate the non-diagonal H$_{ij}$ values.\cite{Am78} All valence electrons were taken into account in the calculations and the basis set consisted of Slater-type orbitals of double-$\zeta  $ quality for C $2s$ and $2p$, S $3s$ and $3p$ and of single-$\zeta $ quality for H. The ionization potentials, contraction coefficients and exponents were taken from previous works.\cite{Ve94,Vi07b}

Alternating current
(1 $\mu$A, 77 Hz) and (5 to 10 $\mu$A, 20 kHz) was injected
parallel to the normal to the conducting plane for measurements of the
interlayer zero-field resistance and magnetoresistance,
respectively. Electrical contacts to the crystal were made using annealed platinum wires of 20
$\mu$m in diameter glued with graphite paste. Hydrostatic pressure was applied up to 1.0 GPa in an
anvil cell designed for isothermal measurements in pulsed magnetic
fields.\cite{Na01} In the following, the pressure applied at room
temperature is considered.
Magnetoresistance experiments were performed up to 55 T in pulsed
magnetic field with pulse decay duration of 0.36 s, in the
temperature range from 1.5 K to 4.2 K. Magnetic field was applied
normal to the conducting plane. A lock-in amplifier
with a time constant of 30 $\mu$s was used to detect the signal
across the potential contacts. Analysis of the oscillatory
magnetoresistance is based on discrete Fourier transforms of the
data, calculated with a Blackman window.

\section{\label{sec:zero} Zero-field properties}

\begin{figure}
\includegraphics[width=0.75\columnwidth,clip,angle=0]{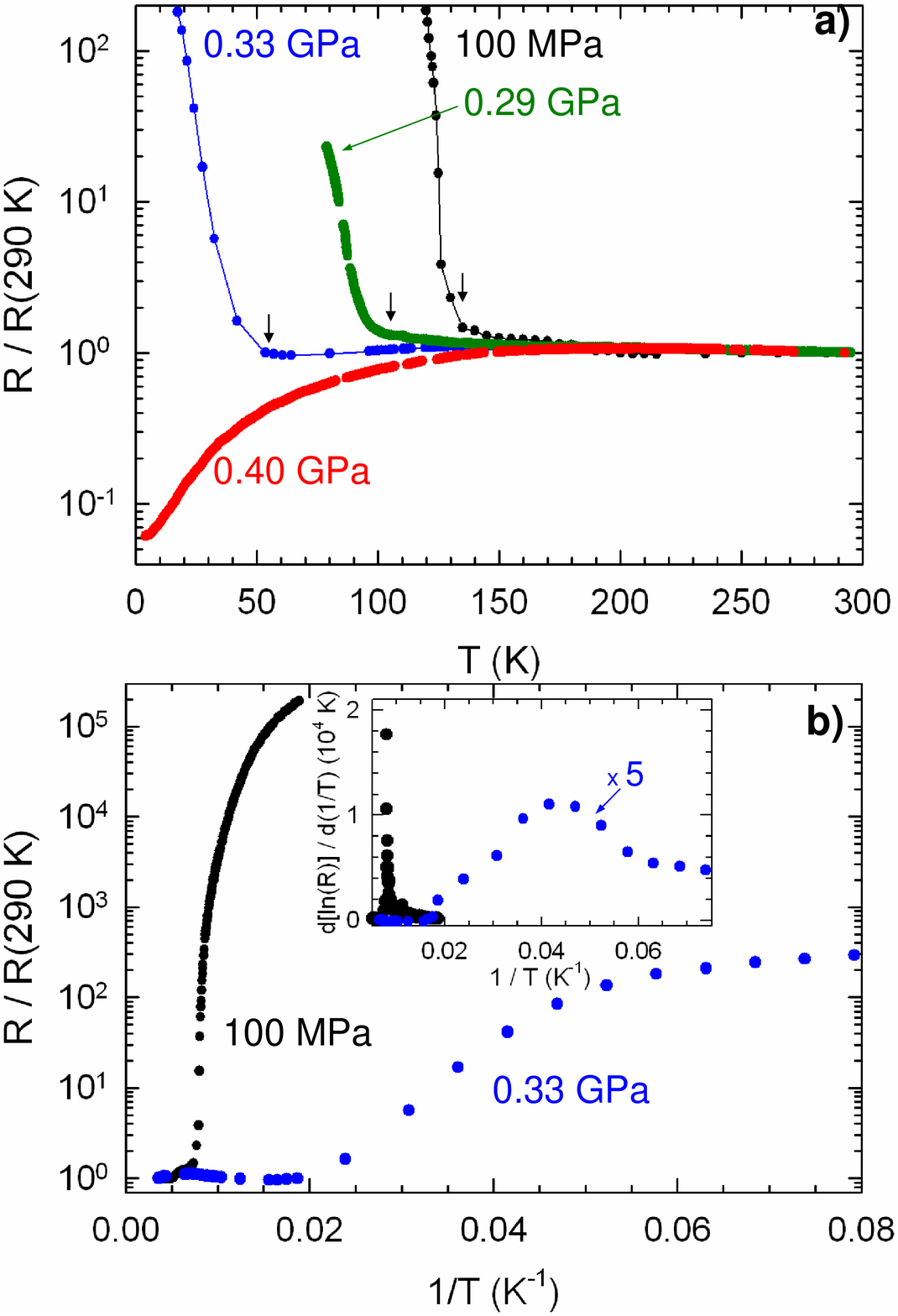}
\caption{\label{Fig:RT} (color online) (a) Temperature dependence of the zero-field resistance at few applied pressures. Arrows indicate the onset of the resistance rise on cooling (T$_{onset}$). (b) Arrhenius plot of the resistance at ambient pressure and 0.33 GPa. The maximum of the logarithmic derivative displayed in the inset yields the temperature T$_c$.}
\end{figure}

\begin{figure}
\includegraphics[width=0.75\columnwidth,clip,angle=0]{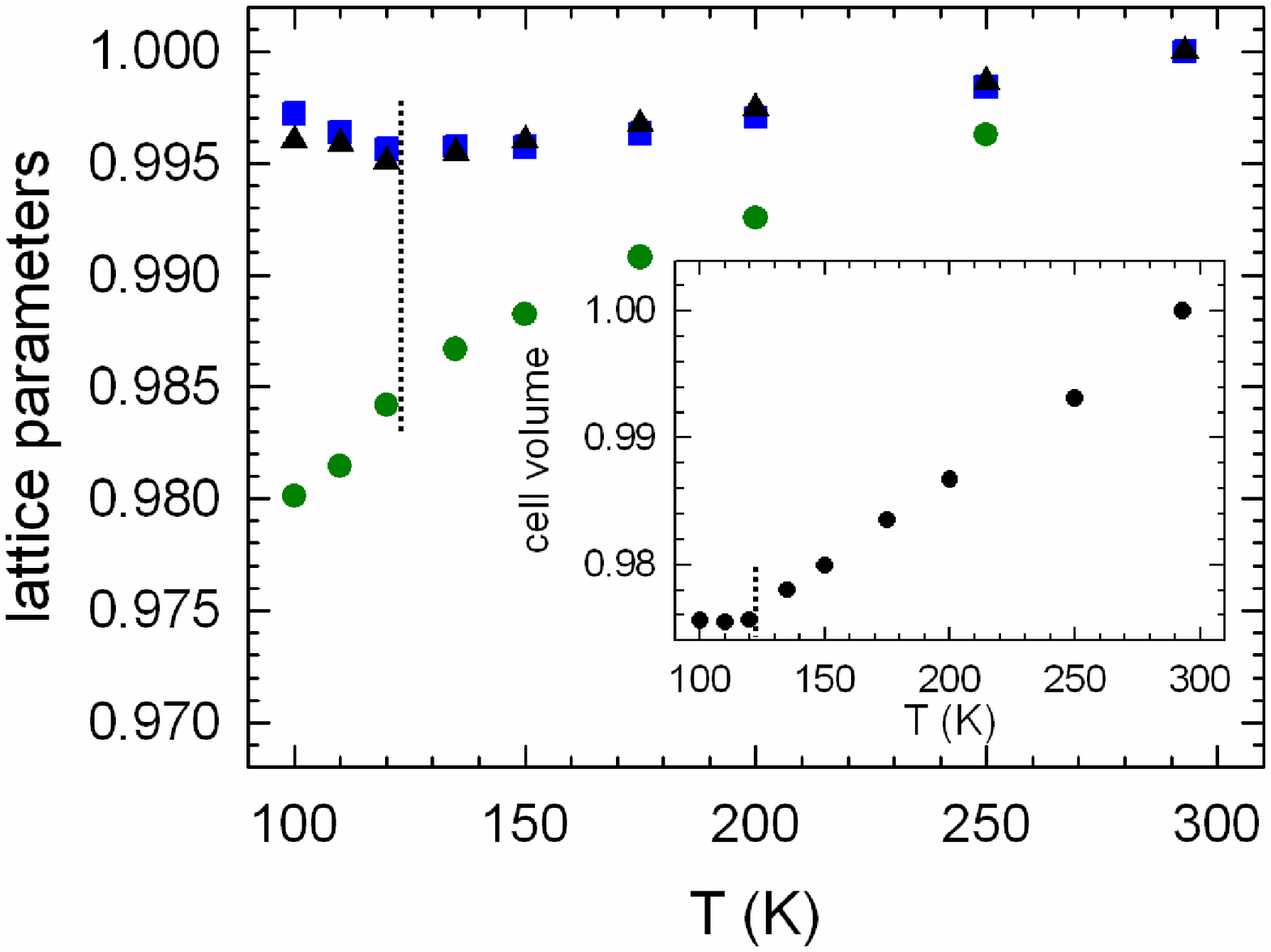}
\caption{\label{Fig:LatticeParam} (color online) Temperature dependence of lattice parameters $a$ (circles), $b$ (squares) and $c$ (triangles). Data are normalized to their value at 293 K. Corresponding data for unit cell volume is displayed in the insert. Vertical lines mark T$_c$ = 122 K deduced from zero-field resistance measurements at ambient pressure of Fig.~\ref{Fig:RT}}
\end{figure}

At ambient pressure, a strong increase of the zero-field resistance is observed in Fig.~\ref{Fig:RT}(a) as the temperature is lowered. No hysteresis is detected upon heating. Arrhenius plot of this data exhibits a clear inflection point as reported in Fig.~\ref{Fig:RT}(b). The metal-insulator transition temperature (T$_c$ = 122 $\pm$ 5 K) can therefore be obtained through the sharp maximum of the parameter d[ln(R)]/d(1/T). More insight on this behavior can be obtained through X-ray diffraction data. (BEDT-TTF)$_8$[Hg$_4$Br$_{12}$(C$_6$H$_5$Br)$_2$] crystallizes in the triclinic
crystal system, space group \textit{P}$\overline{1}$. Upon lowering the temperature down to 100 K, no symmetry change is observed. However, the temperature dependence of the cell parameters and volume reported in Fig.~\ref{Fig:LatticeParam} exhibit a kink at T$_c$ which indicates that the transition is linked to a structural change.

\begin{figure}
\includegraphics[width=0.75\columnwidth,clip,angle=0]{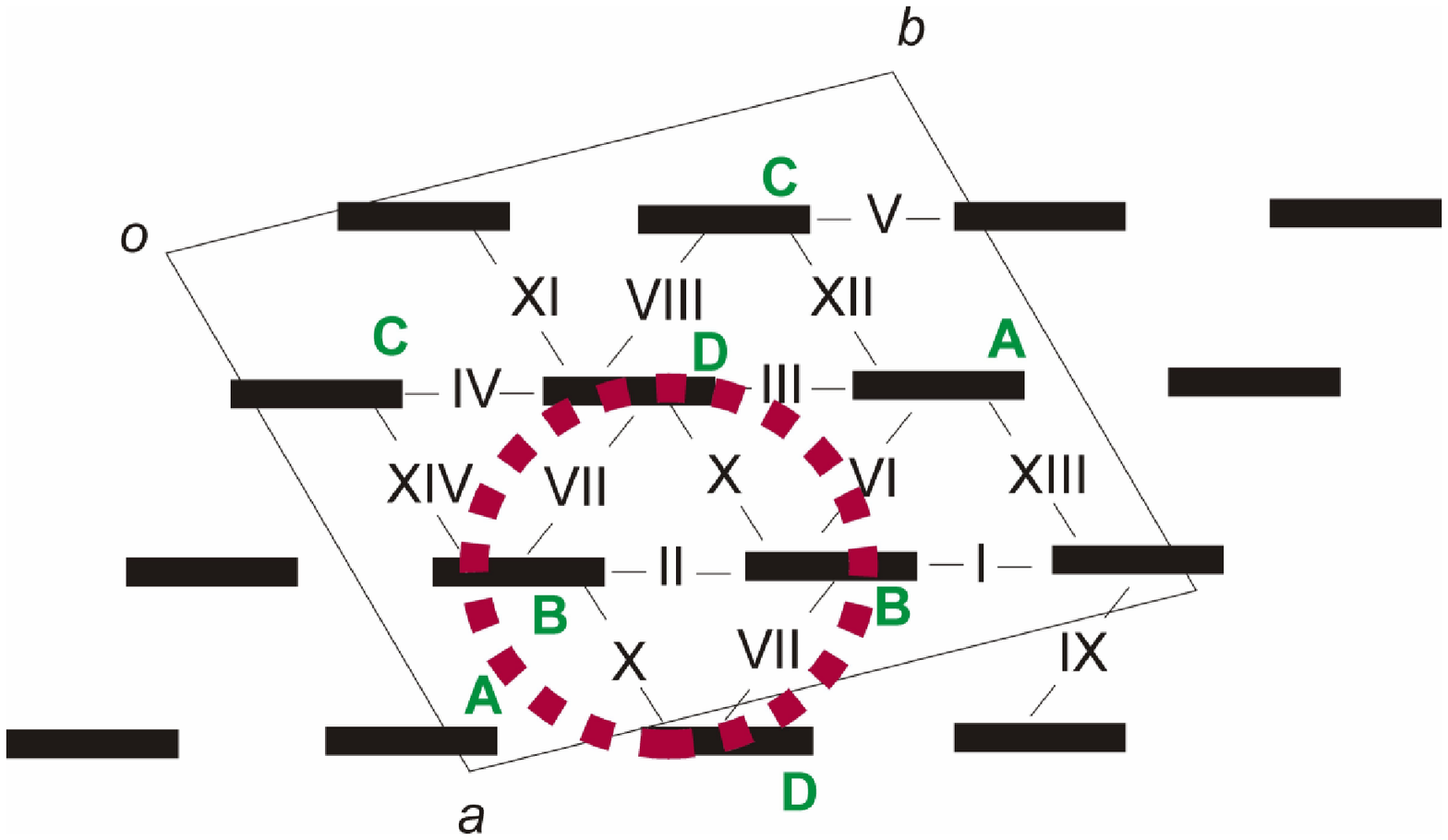}
\caption{\label{Fig:Fig_4} (color online) Donor lattice of (BEDT-TTF)$_8$[Hg$_4$Br$_{12}$(C$_6$H$_5$Br)$_2$] where capital letters and roman numerals label the different types of donors and intermolecular interactions, respectively. The ellipse in dotted line highlights the part of the lattice where main structural changes should occur according to the electronic structure data. }
\end{figure}

To gain some insight on the origin of this change, the evolution of the crystal structure can be considered as the temperature is lowered from room temperature. However, in view of the large number of structural parameters, associated with the big unit cell (V$_{cell}$ = 3720 ${\AA}^3$ at room temperature), that should account for the relatively small change evidenced in the data of Fig.~\ref{Fig:LatticeParam}, it seems difficult to extract information from a structural study alone. Under such circumstances it is usually more informative to consider first the electronic structure. Because of the directionality of the S and C $p$ orbitals leading to the HOMO of BEDT-TTF, the HOMO...HOMO intermolecular interactions are quite sensitive to small structural variations and their analysis is extremely useful for understanding the relationship between features of the crystal and electronic structures (see e.g. Refs. \onlinecite{Zo09,Zv10}). The so-called $\beta _{HOMO-HOMO}$ intermolecular interaction energies \cite{Wh85} for the different donor...donor interactions as well as the HOMO energies of the different BEDT-TTF donors as a function of temperature are considered in the following. As shown in Fig.~\ref{Fig:Fig_4}, the donor lattice of the present salt contains four different BEDT-TTF molecules and fourteen different intermolecular interactions. Five of the interactions (I to V) are lateral $\pi -\pi $ interactions, four (VI to IX) are interactions within the stacks running along the $(a-b)$ direction and five (X to XIV) are interactions along the step-chains, which are usually the strongest for this type of lattice. The different $\mid \beta _{HOMO-HOMO}\mid $ values and HOMO energies calculated at 200 K, 150 K and 120 K are reported in Table~\ref{tab.beta_Homo}. Let us note that the three crystal structures on which these calculations are based on were determined for the same crystal.

\begin{table}
\caption[Refinement]{\label{tab.beta_Homo}Calculated values of $\mid\beta_{HOMO-HOMO}\mid$ for the fourteen different donor...donor interactions and HOMO energies (eV) of the four different donors of (BEDT-TTF)$_8$[Hg$_4$Br$_{12}$(C$_6$H$_5$Br)$_2$]. Labels are the same as in Fig.~\ref{Fig:Fig_4}. }

  \begin{tabular}{c c c c}
    \hline
    &\multicolumn{2}{c}{$\mid\beta_{HOMO-HOMO}\mid$ (eV)} \tabularnewline
    \hline
    Interaction & 200 K &  150 K & 120 K \tabularnewline
    \hline
    I &	0.1312 & 0.1385 & 0.1383\tabularnewline
    II & 0.1093	& 0.1113 & 0.1000\tabularnewline
    III	& 0.1434 & 0.1560 & 0.1525\tabularnewline
    IV & 0.1328 & 0.1416 & 0.1418\tabularnewline
    V & 0.1197& 0.1290& 0.1222\tabularnewline
    \hline
    VI & 0.1132 & 0.1319 & 0.1345\tabularnewline
    VII & 0.1299 & 0.1061 & 0.1284\tabularnewline
    VIII & 0.0719 & 0.0863 & 0.0847\tabularnewline
    IX & 0.1135 & 0.1158 & 0.1193\tabularnewline
    \hline
    X & 0.2915 & 0.2938 & 0.2861\tabularnewline
    XI & 0.1649 & 0.1565 & 0.1312\tabularnewline
    XII & 0.2257 & 0.2385 & 0.2364\tabularnewline
    XIII & 0.2897 & 0.2928 & 0.2820\tabularnewline
    XIV & 0.3066 & 0.3273 & 0.3242\tabularnewline
    \hline
    \hline
    &\multicolumn{2}{c}{HOMO energy (eV)} \tabularnewline
    \hline
    Donor & 200 K & 150 K & 120 K\tabularnewline
    A & -8.4917 & -8.5048 & -8.5767\tabularnewline
    B & -8.4808 & -8.3775 & -8.4266\tabularnewline
    C & -8.4687 & -8.4443 & -8.4492\tabularnewline
    D & -8.4738 & -8.4561 & -8.6004\tabularnewline
    \hline
  \end{tabular}
\end{table}

Regarding $\mid \beta _{HOMO-HOMO}\mid$ interactions, three different types of behavior can be distinguished in Table~\ref{tab.beta_Homo}: (i) interactions like IX which exhibit a monotonous gradual change, (ii) interactions like I, III, IV, etc. which gradually change to reach a saturation around 150 K, and (iii) interactions like II, VII, X (and to a lesser extent XIII) exhibiting a large non-monotonic change, in particular at low temperature. Whereas behaviors (i) and (ii) are  observed in most organic conductors under thermal contraction, behavior (iii) is \textit {atypical} and certainly points out to some abrupt change at low temperature. According to Fig.~\ref{Fig:Fig_4}, the three interactions II, VII and X  which clearly exhibit this atypical change are all clustered in the region of interaction between donors B and D. Besides, contrary to donors A and C, the temperature dependence of the HOMO energy of B and D is strongly non-monotonic as well. Therefore, it is clear that some abrupt change affecting the electronic parameters occurs in the region highlighted in Fig.~\ref{Fig:Fig_4} around the four donors of type B and D.

\begin{figure*}
\includegraphics[width=1.0\columnwidth,clip,angle=0]{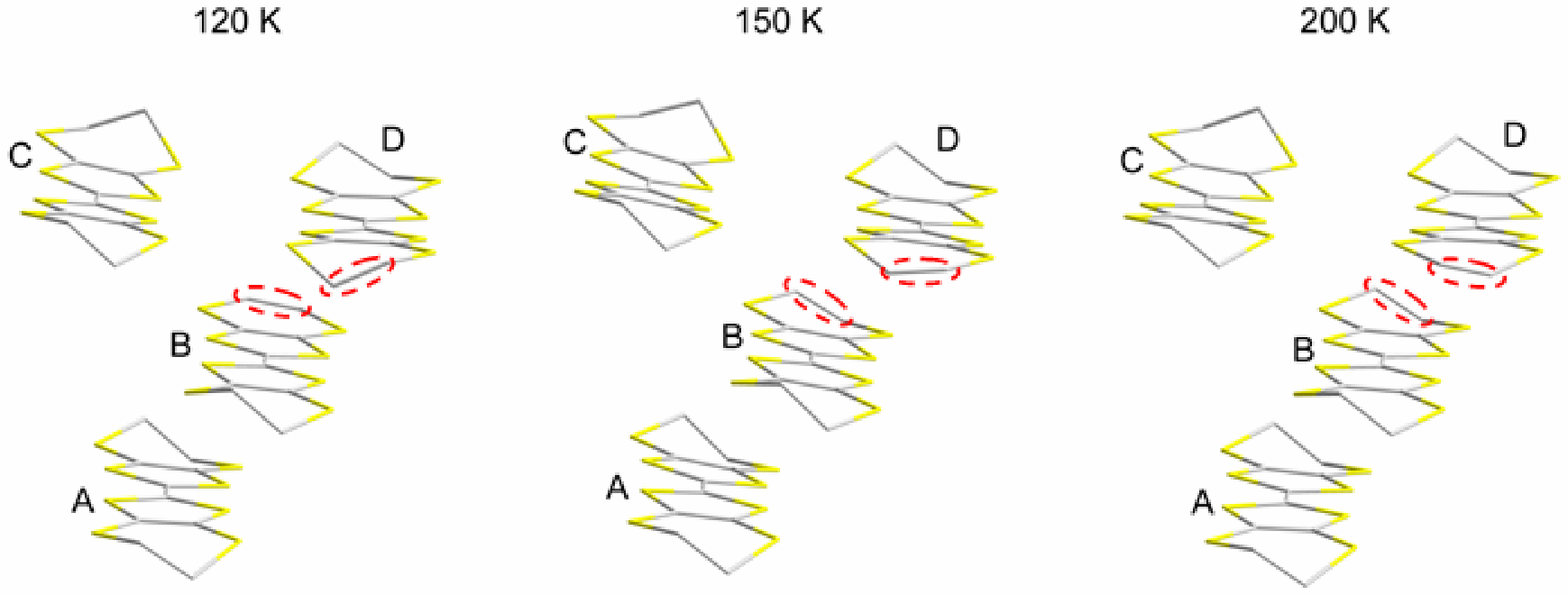}
\caption{\label{Fig:Fig_5} (color online) Evolution of the BEDT-TTF molecules conformation as a function of temperature. Labels are the same as in Fig. \ref{Fig:Fig_4}. Ellipses in dotted lines mark the C-C bonds that experience the most significant changes as the temperature varies (see text). }
\end{figure*}

Taking advantage of these results, we can examine the evolution of the crystal structure as the temperature is varied. At variance with the other structural parameters, the temperature dependence of the outer C-C single bond length of the outer six-membered rings of donors B and D is large and non-monotonic. Most interestingly, these features are related with a conformational change affecting these six-membered rings which change as displayed in Fig.~\ref{Fig:Fig_5}. The same change was also observed in another set of structural refinements for a different crystal. Consequently, with both the crystallographic and electronic structure results in mind we can quite firmly conclude that the structural change occurring around 122 K is related to a conformational change of two of the BEDT-TTF donors and leads to the kink in Fig.~\ref{Fig:LatticeParam}.

According to band structure calculations reported in Fig.~\ref{Fig:BS_FS}, a metallic state is obtained in the high temperature range. At room temperature and 200 K, the FS is composed of one elongated electron orbit with an area  $A_e$=13 \% of the FBZ area and two hole orbits with area $A_h$=$A_e$/2. Although this topology is different from that of (Cl, Br) displayed in Fig. \ref{Fig:BS_FS}(e) it can be regarded as closely related, the only qualitative difference being a splitting of the hole orbit. In particular, we are still dealing with a compensated metal. As the temperature is lowered, a sizeable shrinkage of the orbits area is observed since, at 150 K, the electron orbits are also split in two parts with an area which reduces to only 1 \% of the FBZ area each. Further temperature lowering finally leads to gap opening (E$_g$ = 38 meV at 120 K, see Fig.~\ref{Fig:BS_FS}(d)) and to the orbits suppression, accounting for the metal-insulator transition observed in the ambient pressure data of Fig. \ref{Fig:RT}. The difference between the behavior of (Cl, Br) and (Br, Br) can be accounted for by chemical pressure. Indeed, since Br atoms entering the HgX$_3$ anions are larger than Cl atoms, the distance between BEDT-TTF molecules is larger for X = Br than for X = Cl. As a result, band overlapping is less pronounced in (Br, Br), leading to a less conducting compound and leaving enough room for molecular changes or sliding as the temperature decreases. Coming back to the origin of the metal-to-insulator transition, we conclude that thermal contraction prepares the system by considerably reducing the bands overlap. At T$_c$, this overlap is already quite small and a sudden conformational change of half of the donor molecules lead to the band gap opening, hence to the abrupt change in conductivity regime reported in Fig.~\ref{Fig:RT}.


\begin{figure}
\includegraphics[width=.75\columnwidth,clip,angle=0]{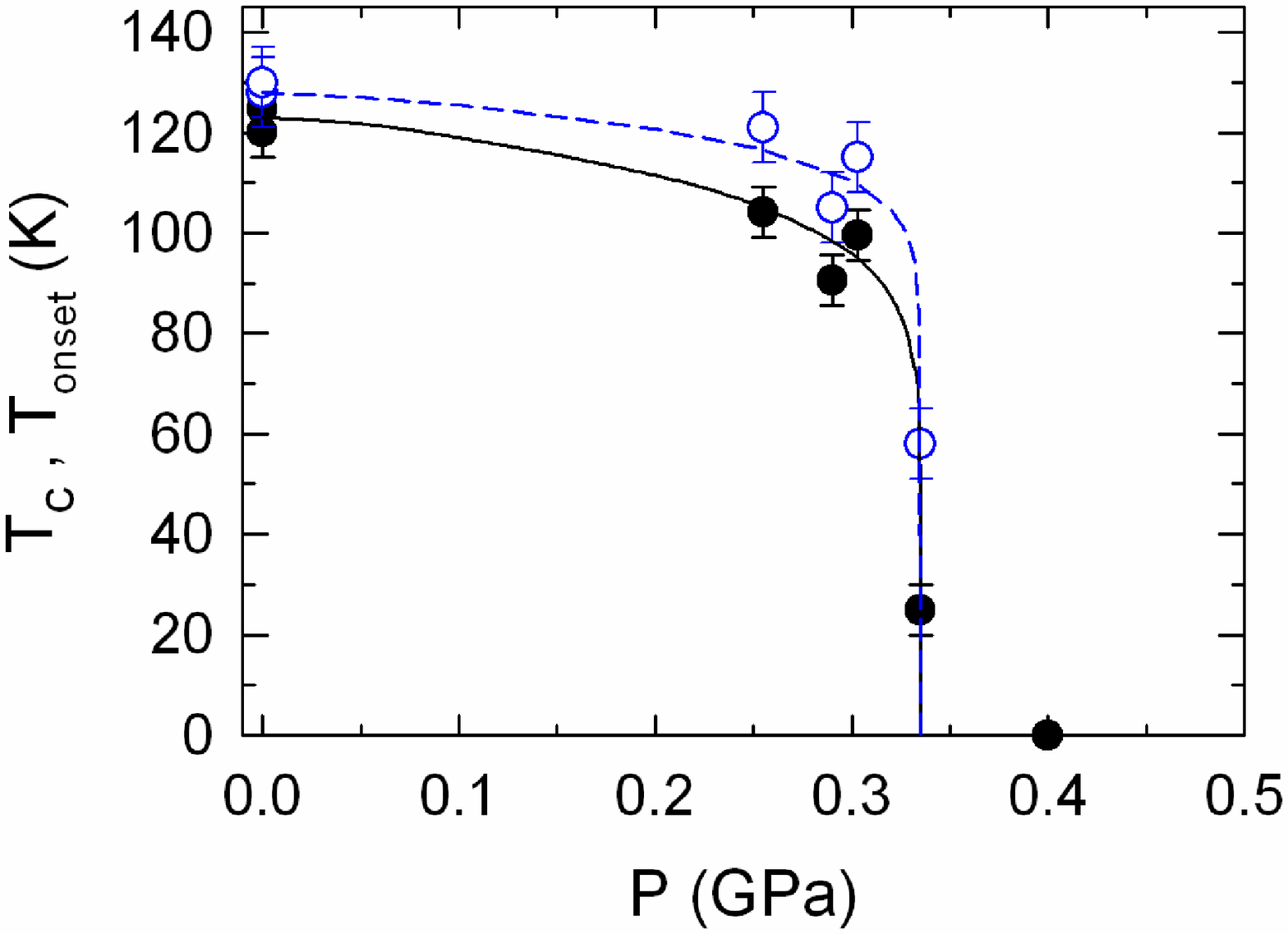}
\caption{\label{Fig:Tc(P)} (color online) Applied pressure dependence of  T$_c$ (solid symbols) and T$_{onset}$ (open symbols) as defined in Fig.~\ref{Fig:RT}. Lines are guides to the eye.}
\end{figure}

\begin{figure}
\includegraphics[width=1\columnwidth,clip,angle=0]{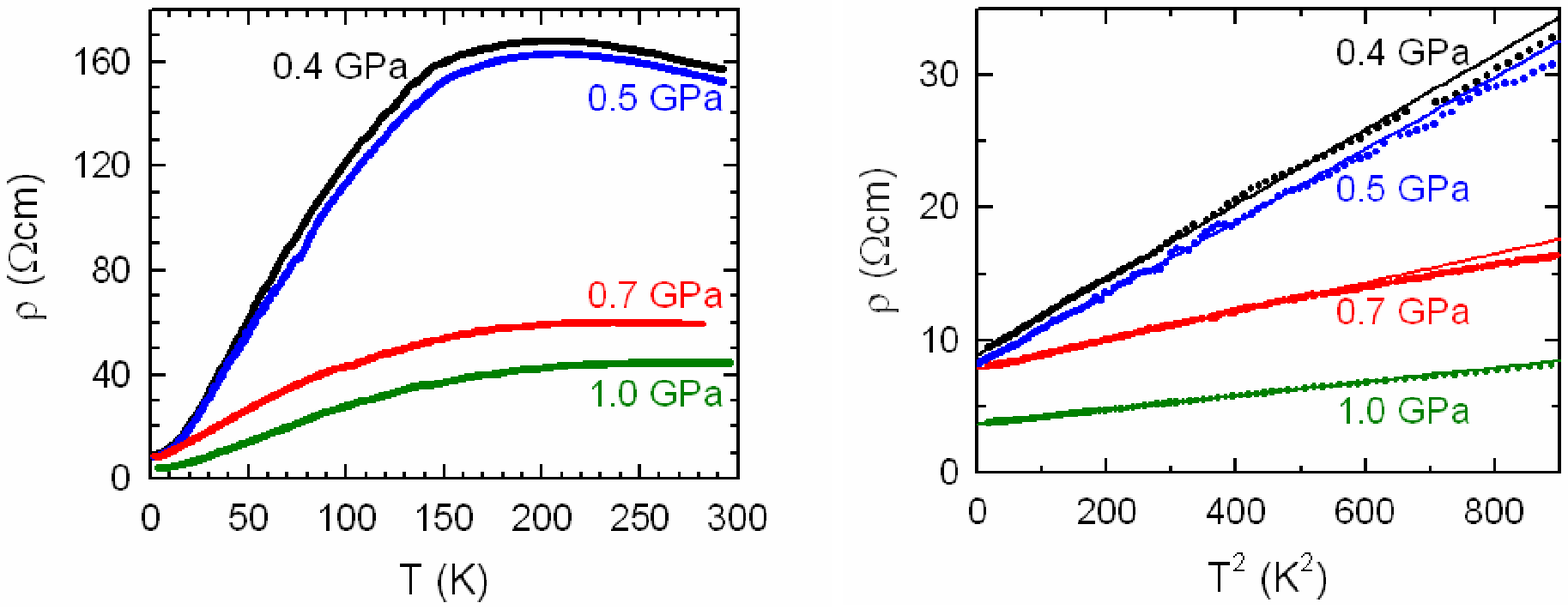}
\caption{\label{Fig:RT_metal} (color online) (a) Temperature dependence of the resistivity in the pressure-induced metallic state above P$_c$ and (b) low temperature part of the data plotted as a function of T$^2$.}
\end{figure}

As the applied pressure increases, T$_c$ decreases, the transition broadens and the ground state becomes clearly not insulating at pressures of a few tenth of a GPa, as displayed in Fig.~\ref{Fig:RT}(b). As a result, the temperature at which the resistance start to increase on cooling (T$_{onset}$) is significantly higher than T$_c$ in this pressure range.  Nevertheless, both T$_c$ and T$_{onset}$ go to zero at P$_c$ = 0.34 $\pm$ 0.02 GPa as observed in Fig.~\ref{Fig:Tc(P)}.

The temperature dependence of the resistivity in the pressure-induced metallic state is very similar to the
data obtained for (Cl, Br).\cite{Vi08b} More precisely, a bump is
observed followed by a metallic behavior (see Fig.
\ref{Fig:RT_metal}(a)). This behavior, observed in many correlated organic conductors, is attributed to a combined influence of disorder and electron correlations.\cite{St99,Me00,Li03,St05,Vi08b} Below about 25 K, the resistivity follows
a T$^2$ variation ($\rho$ = $\rho_0$ + $A$T$^2$, see Fig.
\ref{Fig:RT_metal}(b)) which is generally regarded as a signature of a
correlated Fermi liquid. As already observed
for the (Cl, Br) compound, the $A$ coefficient strongly
decreases as the applied pressure increases, suggesting a reduction of many body strength under pressure.\cite{Vi08b}

\section{\label{sec:field} Field-dependent properties under pressure}

\begin{figure}
\includegraphics[width=0.75\columnwidth,clip,angle=0]{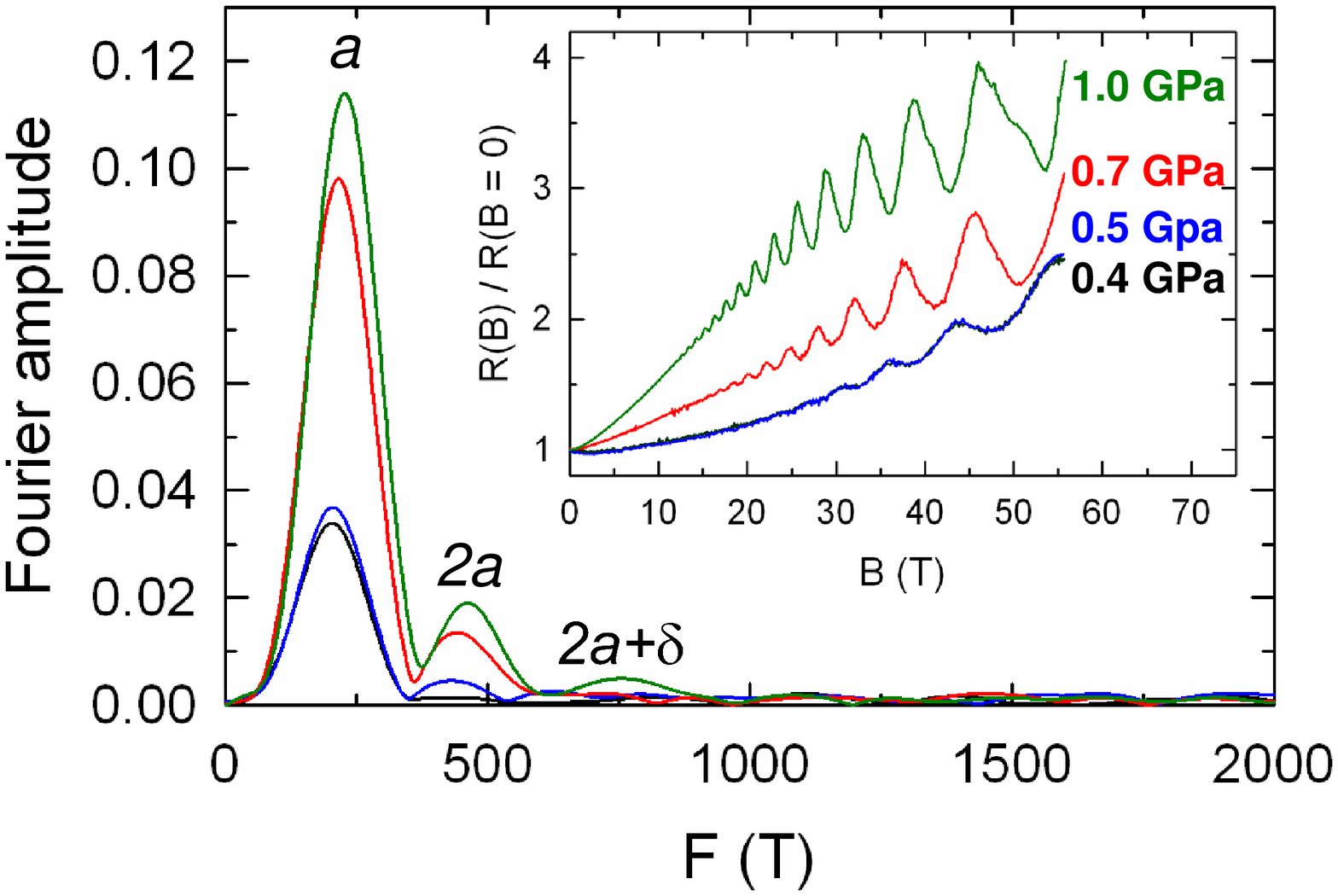}
\caption{\label{Fig:R(B)_TF} (color online) Fourier analysis of the magnetoresistance data at 1.5 K displayed in the insert. Labels refer to the Fermi surface displayed in Fig. \ref{Fig:BS_FS}.}
\end{figure}

\begin{figure}
\includegraphics[width=0.75\columnwidth,clip,angle=0]{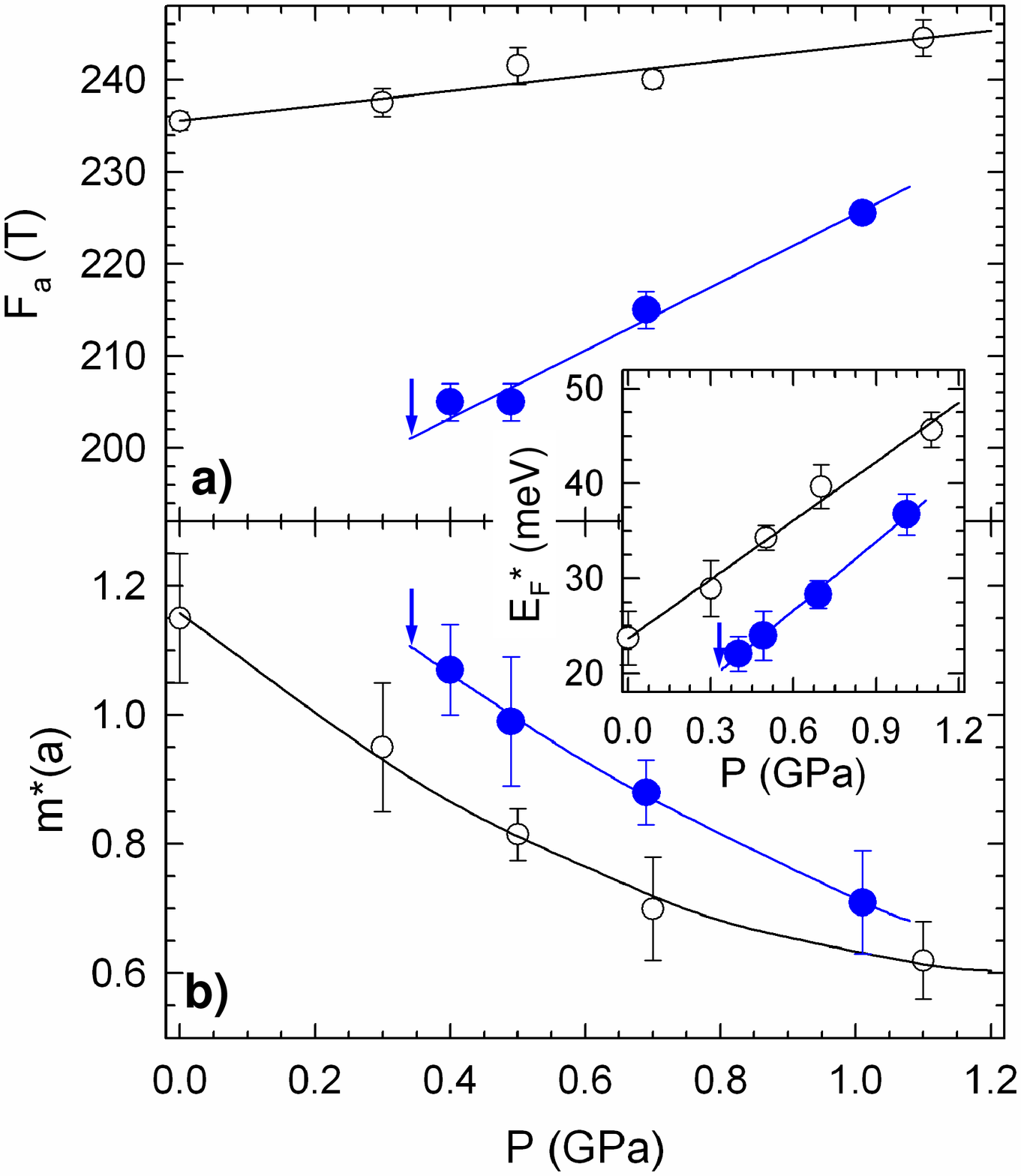}
\caption{\label{Fig:F_mc_EF(P)} (color online) Applied pressure dependence of (a) the oscillations frequency F$_a$ and (b) the effective mass m$^*$($a$). The Fermi energy calculated as E$^*_F$ = e$\hbar$F$_a$/m$^*$($a$) is displayed in the insert. Solid and open symbols stand for (Br, Br) and (Cl, Br), respectively. Arrows marks the critical pressure P$_c$. Data for (Cl, Br) are from Ref.~\onlinecite{Vi08b}.}
\end{figure}

As displayed in Fig.~\ref{Fig:R(B)_TF}, SdH oscillations are observed in the pressure-induced metallic state. Fourier analysis reveal a mean peak that ranges from 205 $\pm$ 2 T at 0.4 GPa to 225.5 $\pm$ 1.0 T at 1.0 GPa (see Fig.~\ref{Fig:F_mc_EF(P)}). These values correspond to 9.5 \% and 10.5 \% of the ambient pressure FBZ area at 200K, respectively, which is close to the $a$ orbit area of (Cl, Cl) and (Cl, Br) and to the ambient pressure electron orbit area of (Br, Br) deduced from band structure calculations at 200 K. The pressure sensitivity of this frequency is much larger than for (Cl, Br), namely, d[ln(F$_a$)]/dP = 0.20 GPa$^{-1}$ and 0.04 GPa$^{-1}$ for (Br, Br) and (Cl, Br),
respectively. In the case of (Cl, Br), the frequency linked to the quantum interferometer $b$, the area of
which is just equal to the FBZ area (F$_b$ = 2F$_a$+F$_{\delta}$+F$_{\Delta}$), is detected in the spectra. Since the same pressure dependence is observed for F$_b$ and F$_a$, it can be concluded that the pressure dependence of F$_a$ is the same as that of the FBZ area. In the case where this statement holds for (Br, Br) it indicates that the FBZ area, hence the unit cell parameters of (Br, Br), are more strongly pressure sensitive. Effect of chemical pressure can also be invoked to account for this difference in pressure sensitivity.

Two other points must be noticed. First, band structure calculations relevant to the metallic state at ambient pressure in the temperature range above T$_c$ predict split hole orbits (see Fig. \ref{Fig:BS_FS}), hence a frequency F$_a$/2. Secondly, oscillatory spectra of Fig.~\ref{Fig:R(B)_TF} contain much less Fourier components than in the case of (Cl, Cl) and (Cl, Br). Indeed, apart the frequency F$_a$ and its second harmonics, only a frequency at 770 $\pm$ 20 T is observed at 1 GPa. Regarding the first point, it can be remarked that, due to the rather small field range in which oscillations are observed, Fourier peaks are rather broad which makes necessary to check if they actually correspond to a single frequency (see e.g. Ref. \onlinecite{Au09}).

\begin{figure}
\includegraphics[width=0.75\columnwidth,clip,angle=0]{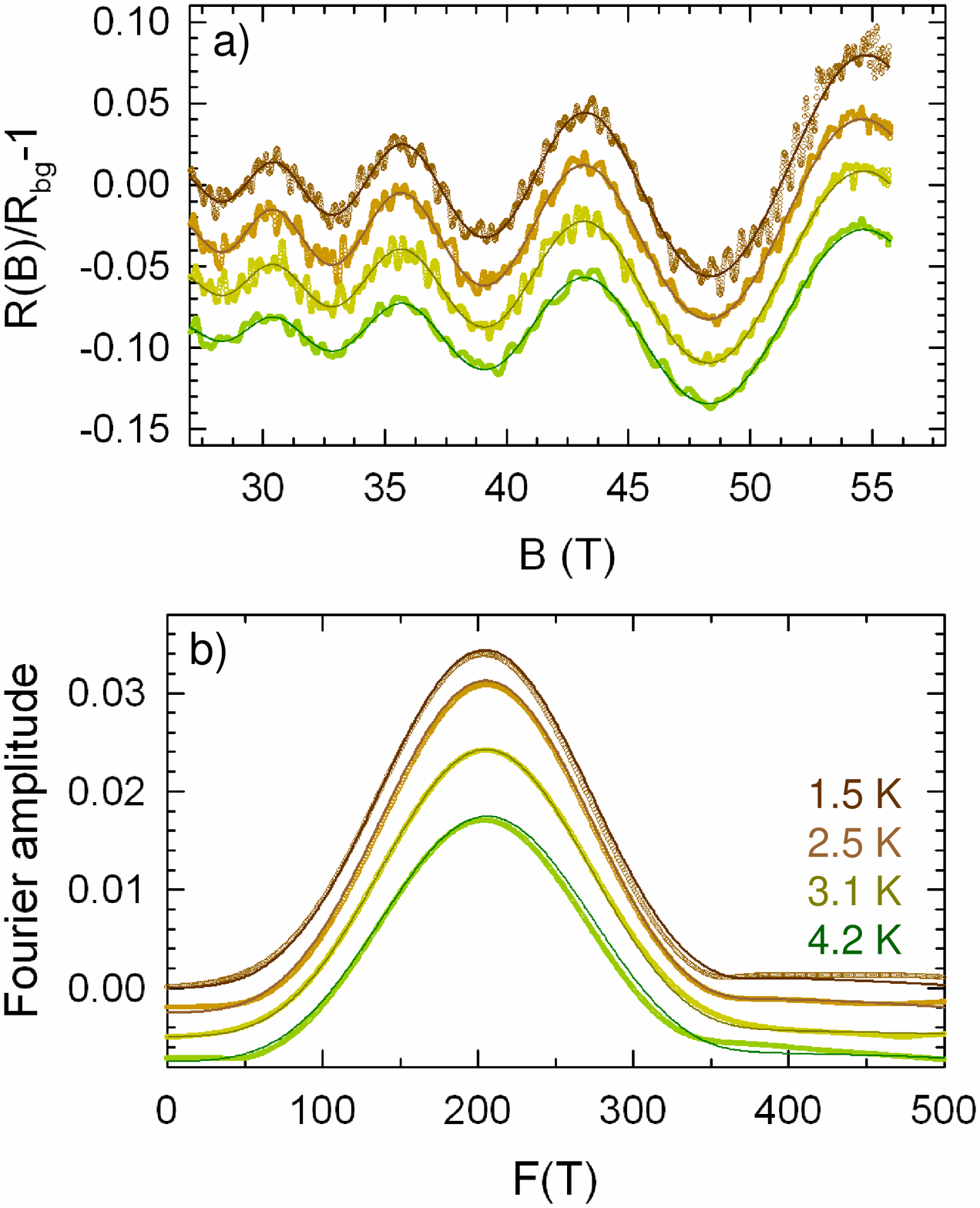}
\caption{\label{Fig:fit_s_TF} (color online) (a) oscillatory magnetoresistance data at 0.4 GPa and best fits of the Lifshits-Kosevich formalism obtained with F$_a$ = 205 T, m$^*$(a) = 1.04 and T$_D$ = 7 K. (b) corresponding Fourier analysis. Symbols and solid lines stand for experimental data and fits, respectively. }
\end{figure}

More insight about these two points can be obtained through field and temperature dependence of the $a$ oscillations
amplitude. In the following, this is considered in the framework of the Lifshits-Kosevich
model, the main ingredients of which being the effective mass (m$^*$), the Dingle temperature which is linked to the scattering rate (T$_D$ = $\hbar$/2$\pi$k$_B\tau$) and the MB field (B$_{0}$).

Fig.~\ref{Fig:fit_s_TF} displays oscillatory magnetoresistance data at 0.4 GPa, which is the studied applied pressure closest to P$_c$, and best fits to the data assuming only one Fourier component is involved. A good agreement between fits and experimental data is observed for both the oscillatory magnetoresistance and the corresponding Fourier analysis. This result demonstrates that no Fourier component with frequency F$_a$/2 enters the oscillatory spectrum. The data are therefore more in line with a Fermi surface such as that of (Cl, Cl) and (Cl, Br) i.e. composed of one electron and one hole compensated orbit.

As reported in Fig. \ref{Fig:F_mc_EF(P)}(b), m$^*$($a$) strongly increases as the applied pressure decreases, as it is observed for (Cl, Br). In this latter case, the $A$ coefficient of the zero-field resistivity is proportional to [m$^*$($a$)]$^2$ which has been interpreted on the basis of a pressure-induced decrease of electron correlations.\cite{Vi08b,Me00} Although only a very rough agreement with this proportionality is observed for (Br, Br), $A$ strongly decreases as the applied pressure decreases, as observed in Fig.~\ref{Fig:RT_metal}(b). This feature suggests that applied pressure induces a decrease of the electron correlations, as well.

\begin{figure}
\includegraphics[width=0.75\columnwidth,clip,angle=0]{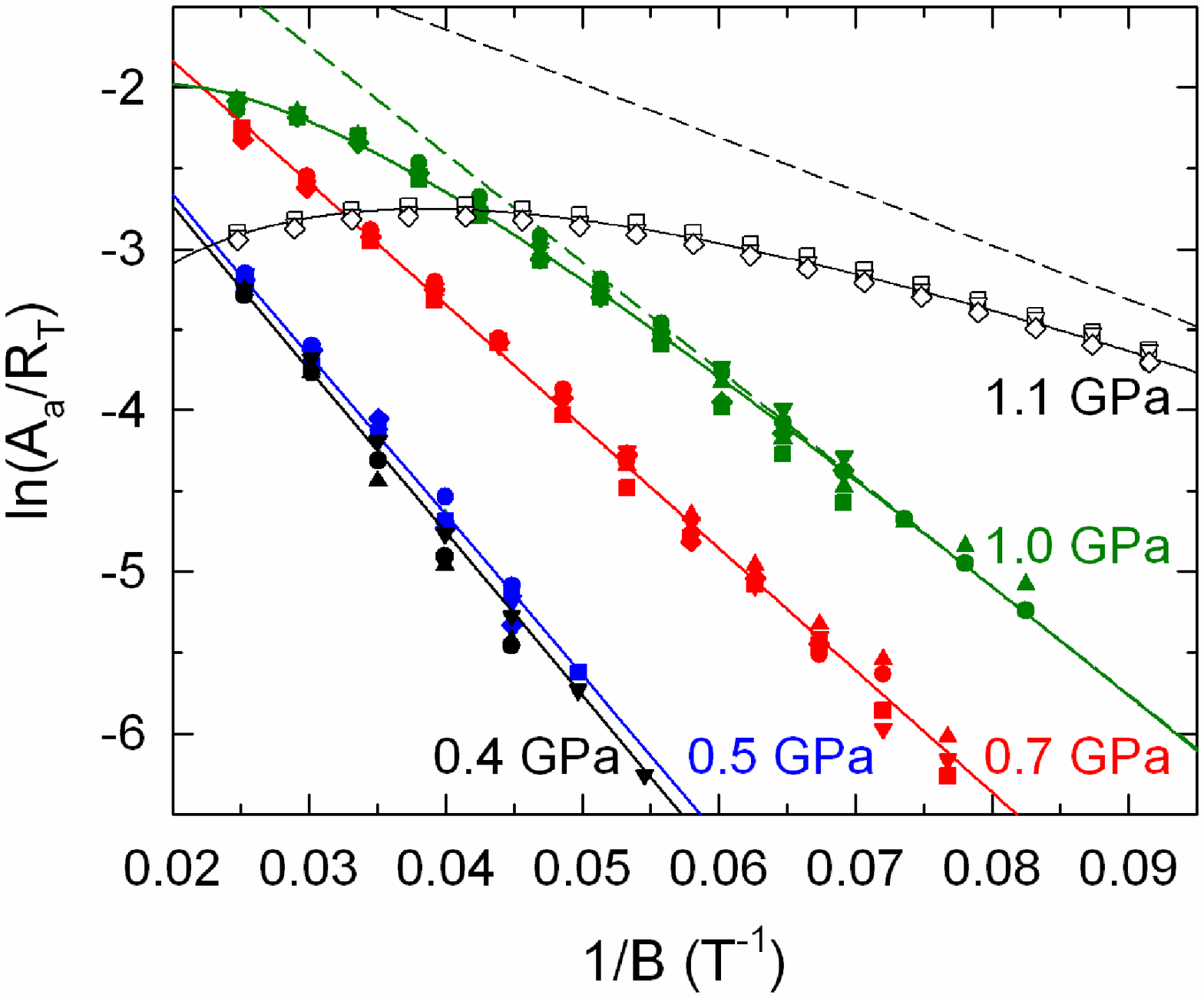}
\caption{\label{Fig:Dingle} (color online) Dingle plots of the $a$ oscillations. Thermal damping factors (R$_T$) are calculated with effective masses reported in Fig.~\ref{Fig:F_mc_EF(P)}. Solid and empty symbols are experimental data of (Br, Br) and (Cl, Br), respectively, for various temperatures in the range 1.5 to 4.2 K. Solid and dashed lines are best fits of the Lifshits-Kosevich formalism to the data, including and excluding the contribution of magnetic breakdown, respectively. Magnetic breakdown field (B$_0$) is infinitely large for (Br, Br) below 1 GPa (i.e. R$_{MB}$=1), while B$_0$ = 50 T and 21 T for (Br, Br) at 1.0 GPa and (Cl, Br) at 1.1 GPa \cite{Vi08b}, respectively. }
\end{figure}

In the pressure range from 0.4 GPa to 0.7 GPa, Dingle plots of Fig. \ref{Fig:Dingle} exhibit a linear behavior which allow to deduce Dingle temperatures in the range 6 to 7 K. In contrast, a clear curvature is observed at 1.0 GPa. This curvature which is concomitant with the appearance of the frequency at 770 $\pm$ 20 T in the spectra, can be accounted for by the decrease of the MB field, hence of the field-dependent MB damping factor which includes four Bragg reflections  for the $a$ oscillations (R$_{MB}$ = [1-exp(-B$_0$/B)]$^2$).\cite{Vi03} Within this assumption, the frequency at 770 T might correspond to the MB orbit $2a$+$\delta$, in which case the frequency linked to $\delta$ amounts to F$_{\delta}$ = 319 $\pm$ 22 T which is very large compared to the data of (Cl, Br). \footnote{Recall that this latter frequency corresponds to an 'orbit' which is forbidden within the semiclassical model of coupled orbits networks although it is observed in (Cl, Cl) and (Cl, Br).\cite{Pr02,Vi03,Au05}} Nonetheless, the effective mass linked to this frequency is 1.3 $\pm$ 0.2 which is close to 2m$^*$($a$), as expected for m$^*$($2a+\delta$).\cite{Vi03} Unfortunately, reliable value of the MB field cannot be derived from the data due to the large Dingle temperature.\footnote{In the case where Bragg reflections are involved in a MB orbit, the parameter A$_a$/R$_T$ goes to
a maximum at a field  B$_{max}$ which should be lower than the maximum field reached in the experiment in order to derive reliable B$_0$ value. For the $a$ oscillations that involves four Bragg reflections, B$_{max}$ = B$_0$/ln(1 + B$_0$e$\hbar$/2$\pi^2$k$_B$m$_e$m$^*_a$T$_D$) \cite{Vi07} which leads to B$_{max}$ as high as 56 T for the data at 1.0 GPa.} Nevertheless, it can be inferred that the MB field decreases as the applied pressure increases, as already observed for (Cl, Br). However, the MB field is much larger in (Br, Br) than in (Cl, Br) at a given applied pressure (see \cite{Vi08b} and open symbols in Fig. \ref{Fig:Dingle}) which justifies the absence of frequency mixing in (Br, Br).

Finally, as it is the case for (Cl, Br), the effective mass strongly decreases as the applied pressure increases. Such a variation, which is much too large to be accounted for by band mass decrease under pressure \cite{Vi08b,Me00}, can be attributed to a pressure-induced decrease of electron correlations since the $A$ coefficient of the T$^2$ law of the zero-field resistivity also strongly decreases under pressure. Phase diagram of strongly correlated systems, in which insulating, magnetic, superconducting and non-Fermi liquids states are observed, also exhibits large overlinear increase or even divergence of the effective mass as approaching the metal-insulator transition. As pointed out in Ref. \onlinecite{Si10}, this behavior can be more clearly evidenced considering an effective Fermi energy  E$^*_F$ = e$\hbar$F$_a$/m$^*$($a$) which includes both the effective mass, eventually renormalized by interactions, and the electronic density through the oscillation frequency. As examples, this parameter goes to zero at specific doping values in underdoped superconducting cuprates for which quantum critical points are suggested \cite{Si10} and at the superconducting-SDW transition in iron pnictides BaFe$_2$(As$_{1-x}$P$_x$)$_2$.\cite{Sh10} At variance with the above examples and with numerous strongly correlated fermion systems, the studied compound contains no magnetic atom and no superconducting ground state is observed in the (temperature-pressure) phase diagram, at least down to 1.5 K. As evidenced in Fig.~\ref{Fig:F_mc_EF(P)}, despite the large pressure sensitivity of the effective mass, E$^*_F$ remains finite at P$_c$ which is in line with a Fermi liquid picture.

\section{\label{sec:discussion}Summary and conclusion}

According to band structure calculations relevant to the ambient pressure state in the high temperature range, the Fermi surface of (Br, Br) is composed of one electron orbit with area A$_e$ = 13 \% of the FBZ area and two hole orbits with area A$_e$/2. This topology is closely related to that of (Cl, Cl) and (Cl, Br), although the hole orbit is not split in the latter cases. As the temperature decreases, the orbits area strongly decreases leading to gap opening below T$_c$ = 122 K, in line with the temperature dependence of the resistance. This behavior, which is not observed in (Cl, Cl) and (Cl, Br) can be attributed to a chemical pressure effect in which Br atoms of the [HgBr$_3$]$^-$ anions, larger than Cl atoms, reduce the compression of donors molecules. As a result, band overlapping is reduced which accounts for the different FS topology at room temperature. Besides, as the temperature is lowered, a change of two of the BEDT-TTF molecules conformation and a non-monotonous temperature dependence of the lattice parameters around T$_c$ are observed.

A pressure-induced metallic ground state is observed above P$_c$ = 0.34 GPa. SdH oscillations spectra observed in the pressure range between 0.4 GPa and 0.7 GPa exhibit only one Fourier component and its second harmonic. Frequencies are in agreement with both band structure calculations in the high temperature range for the electron orbits of (Br, Br) and data for the compensated orbits of (Cl, Cl) and (Cl, Br). No Fourier component with frequency F$_a$/2 is observed which suggests that the FS under pressure is similar to that  of (Cl, Cl) and (Cl, Br). The large pressure sensitivity of F$_a$, and likely of the FBZ area, can be ascribed to the reduced chemical pressure as well.

Effective mass and coefficient of the T$^2$ law of the zero-field resistivity strongly decrease as the applied pressure increases. As in the case of (Cl, Br) this can be attributed to a pressure-induced decrease of many body effects. Nevertheless, at variance with more strongly correlated systems for which a superconducting state is generally observed in the vicinity of the insulating state in phase diagram, the effective Fermi energy remain finite at P$_c$ which is in line with a Fermi liquid picture throughout the phase diagram of the studied charge transfer salt family.

\begin{acknowledgments}
This work has been supported by FP7 I3 EuroMagNET II, DGI-Spain (Grants No. FIS2009-12721-C04-03 and CSD2007-00041) and RFBR grant 11-03-01039a.
\end{acknowledgments}

\bibliography{BrBr}

\end{document}